\documentclass[12pt]{article}

\pdfoutput=1

\usepackage[super]{cite}     
\usepackage{amsmath} 	
\usepackage{amssymb}  
\usepackage{graphicx} 
\usepackage{upgreek}

\usepackage{sectsty}
\sectionfont{\large}

\newcommand{\nature}{2008_Nature_Thompson_Harris}
\newcommand{\njp}{2008_NJP_Jayich_Harris}

\newcommand{\degrees}{$^\circ$ }
\newcommand{\f}[2] {Fig.\ \ref{fig#1}#2}
\newcommand{\F}[2] {Figure \ref{fig#1}#2}

\begin{document}

\title{Strong and Tunable Nonlinear Optomechanical Coupling in a Low-Loss System}

\author{J. C. Sankey$^1$, C. Yang$^1$, B. M. Zwickl$^1$, A. M. Jayich$^1$, J. G. E. Harris$^{1,2}$}

\maketitle

\begin{abstract}


A major goal in optomechanics is to observe and control quantum behavior in a system consisting of a mechanical resonator coupled to an optical cavity. Work towards this goal has focused on increasing the strength of the coupling between the mechanical and optical degrees of freedom; however, the form of this coupling is crucial in determining which phenomena can be observed in such a system. Here we demonstrate that avoided crossings in the spectrum of an optical cavity containing a flexible dielectric membrane allow us to realize several different forms of the optomechanical coupling. These include cavity detunings that are (to lowest order) linear, quadratic, or quartic in the membrane's displacement, and a cavity finesse that is linear in (or independent of) the membrane's displacement. All these couplings are realized in a single device with extremely low optical loss and can be tuned over a wide range \textit{in situ}; in particular, we find that the quadratic coupling can be increased three orders of magnitude beyond previous devices. As a result of these advances, the device presented here should be capable of demonstrating the quantization of the membrane's mechanical energy.

\end{abstract}


Nearly all optomechanical systems realized to date can be characterized by a linear relationship between the optical cavity's detuning $\omega(x)$ and the displacement of the mechanical element $x$.\cite{2007_OpticsExpress_Kippenberg_Vahala} In the classical regime this ``linear'' optomechanical coupling has enabled powerful laser cooling and sensitive displacement readout of the mechanical element. \cite{2009_nphys_Schliesser_Kippenberg,2009_nphys_Groblacher_Aspelmeyer,2009_nphys_Park_Wang,2006_PRL_Arcizet_Rousseau,2009_NJP_Abbott_Zucker,2009_nphys_Anetsberger_Kippenberg} As $\omega^{\prime} \equiv \partial \omega / \partial x$ increases this linear coupling becomes stronger, and it should become possible to observe quantum effects such as laser-cooling to the mechanical ground state\cite{2007_PRL_Marquardt_Girvin, 2007_PRL_Rae_Kippenberg}, quantum-limited measurements of force and displacement\cite{1975_SPU_Braginskii_Vorontsov, 1980_PRL_Caves}, and the production of squeezed light.\cite{1994_PRA_Tombesi_Vitali} In the quantum regime, however, the \textit{form} of the optomechanical coupling plays a crucial role in determining which phenomena are observable. For example, linear coupling provides a continuous readout of $x$, and so precludes a direct measurement of one of the most striking features associated with the quantum regime: the quantization of the mechanical oscillator's energy.

One device that has demonstrated a nonlinear optomechanical coupling consists of a thin dielectric membrane placed inside a Fabry-Perot cavity.\cite{\nature} With the membrane positioned at a node (or antinode) of the intracavity standing wave, $\omega(x) \propto x^2$ to lowest order. This ``quadratic'' optomechanical coupling is compatible with a quantum nondemolition (QND) readout of the membrane's energy $H_m = \hbar \omega_m n_m$, where $\omega_m$ is the membrane's resonant frequency and $n_m$ is the membrane's phonon number. Two distinct schemes have been proposed for using this quadratic coupling to demonstrate the quantization of the membrane's energy. Both schemes assume the membrane is laser-cooled to mean phonon number $\langle n_m \rangle < 1$. Numerical estimates indicate this level of cooling should be feasible for the device described here, provided it is pre-cooled cryogenically.\cite{2007_PRL_Marquardt_Girvin,2007_PRL_Rae_Kippenberg}

The goal of the first scheme is to monitor $n_m$ with resolution sufficient to observe individual quantum jumps. This is not feasible with the devices demonstrated to date, and would require substantial improvements to the quadratic coupling strength $\omega^{\prime\prime} \equiv \partial^{2} \omega / \partial x^{2}$, the membrane's optical absorption, and the membrane's mechanical properties.\cite{\nature}

In the second scheme, the laser-cooled membrane would be mechanically driven from the ground state to a large-amplitude coherent state with $\langle n_m \rangle \gg 1$. The quadratic coupling would then be used to monitor the membrane's energy with resolution sufficient to resolve fluctuations $\propto \sqrt{\langle n_{m}\rangle}$ corresponding to the shot noise of the membrane's phonons. Detailed calculations\cite{2010_arxiv_Clerk_Harris} show that this second scheme is considerably less demanding than the first, though it still requires substantial improvements to $\omega^{\prime\prime}$ and the membrane's optical absorption. 

Here we demonstrate an optomechanical device in which $\omega^{\prime\prime}$ is increased by at least three orders of magnitude, while the membrane's optical absorption is substantially decreased. The device satisfies the requirements for observing phonon shot noise at a bath temperature $T = 300$ mK, and represents substantial progress toward observing individual quantum jumps. The improvement in $\omega^{\prime\prime}$ is achieved by exploiting the full spectrum of the optical cavity's transverse modes, which exhibits numerous avoided crossings as a function of the membrane's position. These crossings were not considered in previous work, which assumed a one-dimensional model of the cavity and only a single transverse optical mode.\cite{\nature,\njp} 

In addition to increased $\omega^{\prime\prime}$, we demonstrate considerable flexibility within a single device: (i) $\omega^{\prime\prime}$ can be varied \textit{in situ} by adjusting the position and tilt of the membrane; (ii) it is possible to tune $\omega^{\prime\prime}$ to zero, thereby realizing a purely quartic optomechanical coupling $\omega(x) \propto x^4$; (iii) the gradient of the cavity relaxation $\kappa^{\prime} \equiv \partial \kappa / \partial x$ can be tuned over a wide range or set to zero; and (iv) cavity modes with different forms of $\omega(x)$ (e.g., linear and quadratic) can be simultaneously addressed using multiple laser frequencies, allowing for simultaneous laser cooling and QND energy readout. Each type of coupling offers distinct functionality, and together they represent a new set of tools for observing and controlling quantum effects in optomechanical systems. We find that the features in the cavity spectrum responsible for these couplings are reproduced by a straightforward theoretical model, allowing for optimization of future devices. 

\section{Strong purely-quadratic optomechanical coupling}

\begin{figure}
\begin{center}
\includegraphics[width=4.9in]{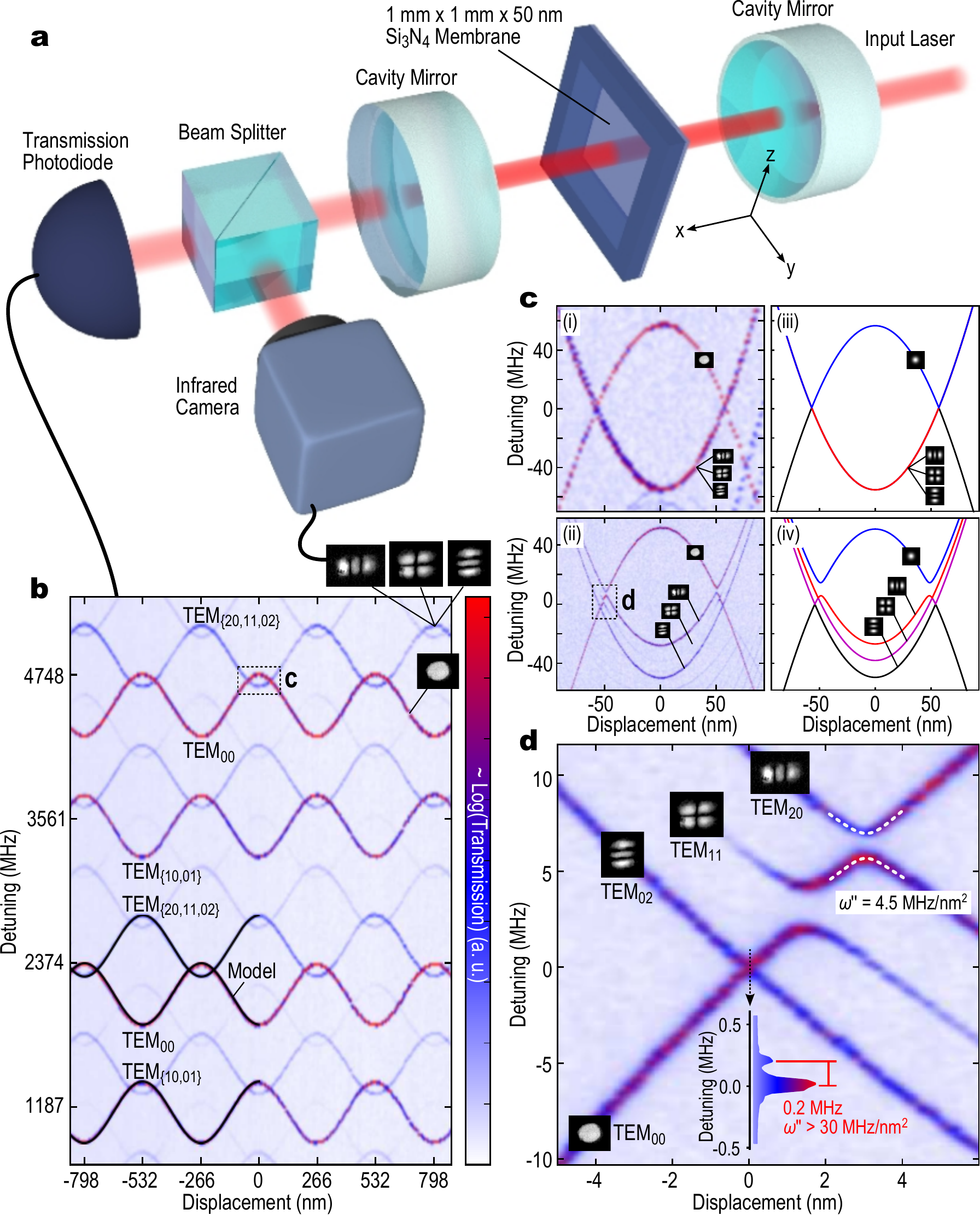}
\end{center}
\caption{Avoided crossings in cavity spectra. \textbf{a} Schematic of our setup. \textbf{b} Cavity transmission as a function of laser detuning and membrane displacement. The membrane is positioned near the cavity waist, and the input laser is coupled to the lower-order transverse cavity modes. The transverse modes corresponding to the strongest transmission peaks are labeled. Images of each of these modes (captured with an infrared camera) are shown. Solid lines show model results. \textbf{c}(i,ii) Refined scans of the box labeled ``c'' in \textbf{a} with the membrane (i) aligned and (ii) tilted by 0.4 mrad. Panels (iii) and (iv) show the corresponding predictions of the model. \textbf{d} Refined scan of the box labeled~``d'' in \textbf{c}(ii).}
\label{fig1}
\end{figure}

Our optomechanical system is shown schematically in \f{1}{a}. Two fixed end mirrors (radius of curvature $R = 5$ cm) form a Fabry-Perot cavity with free spectral range 2.374 GHz (length $L =  6.313$ cm) and empty-cavity finesse $F = 50,000$. A 1-mm-square flexible Si$_3$N$_4$ membrane of nominal thickness $t = 50$ nm and real refractive index Re$[n] = 2.0$ is placed near the cavity waist. The membrane is mounted on a motorized stage, providing control over the membrane's coarse position along the cavity axis ($\tilde{x}$) as well as its tilt about the two transverse axes ($\tilde{y}$ and $\tilde{z}$). Piezoelectric transducers allow for nanometer-scale displacements along $\tilde{x}$.

\F{1}{b} shows the cavity's transmission spectrum as a function of membrane position $x$. The cosine-like detuning curves are similar to those demonstrated previously,\cite{\nature} and achieve $\omega^{\prime\prime} = 30$ kHz/nm$^2$ at their extrema. The data in \f{1}{b} were taken with the laser coupled to several of the cavity's lower-order transverse modes, and a number of apparent crossings between these modes can be seen. We focus specifically on the region highlighted by the dotted box in \f{1}{b}, where the TEM$_{00}$ ``singlet'' and the TEM$_{\{20, 11, 02\}}$ ``triplet'' modes cross. \F{1}{c} shows this region in greater detail with the membrane (i) lying in the $\tilde{y}$-$\tilde{z}$ plane and (ii) tilted by 0.4 mrad about the $\tilde{z}$-axis. The behavior in \f{1}{c}(ii) is ubiquitous among multiplets of higher-order transverse modes: tilting the membrane lifts the multiplet degeneracy, with modes of greater spatial extent along the membrane slope ($\tilde{y}$ in this case) perturbed the most. 

The central result of this article is illustrated in \f{1}{d}, which shows a high-resolution scan of the region indicated by the dotted box in \f{1}{c}. This data demonstrates that the apparent crossings between the various optical modes are avoided, and that at their anticrossings $\omega (x)$ is purely quadratic with $\omega^{\prime\prime}$ substantially greater than 30 kHz/nm$^{2}$. For example, the TEM$_{20}$ - TEM$_{00}$ crossing in \f{1}{d} shows $\omega^{\prime\prime} = 4.5$ MHz/nm$^2$ (dashed white lines). The TEM$_{02}$ - TEM$_{00}$ crossing is not resolved in the main body of \f{1}{d}, but the line-scan in the inset shows a splitting $\approx$ 200 kHz between these modes. Assuming the crossing is of the usual hyperbolic form $\omega (x) = \pm \sqrt{(\omega^{\prime} x)^{2} + \omega_s^{2}}$, where $\omega^{\prime}$ is the slope of $\omega (x)$ far from the crossing and $2 \omega_s$ is the gap at the degeneracy point, the data in the inset indicate $\omega^{\prime\prime} \gtrsim 30$ MHz/nm$^2$, which is three orders of magnitude greater than previously demonstrated.\cite{\nature} 

We note that if the membrane is positioned so that two modes realize a quadratic coupling, other modes will still realize a linear coupling. In \f{1}{d} this occurs at the position $x = 0$ nm: here the eigenmodes formed by the TEM$_{02}$ and TEM$_{00}$ modes exhibit quadratic coupling while the TEM$_{11}$ mode's coupling is linear. This means that lasers tuned to different eigenmodes can simultaneously exploit different forms of the optomechanical coupling.

We can understand the origin of these features by noting that avoided crossings generally reflect a broken symmetry that prevents eigenmodes from becoming degenerate. An ideal empty Fabry-Perot cavity possesses symmetry that allows degeneracy between transverse modes, but in our device we expect this symmetry to be broken for two reasons: the curved wavefronts of the cavity modes may not overlap perfectly with the flat membrane (e.g., if the membrane is tilted or displaced from the cavity waist), and the empty cavity itself may be slightly asymmetric (e.g., owing to imperfect form of the end mirrors). 

To make a quantitative analysis of the cavity spectrum and the features in \f{1}{b-d}, we developed a perturbative solution of the Helmholtz equation to calculate the eigenmodes and eigenfrequencies of a symmetric optical cavity into which a dielectric slab is placed at an arbitrary location and tilt. As described elsewhere\cite{2009_ICAP_Sankey_Harris}, the empty-cavity eigenfrequencies are perturbed by an amount proportional to the eigenvalues of the matrix $\bf{V}$, where $V_{i,j} \propto \iiint\! \psi_{i}(x,y,z) \psi_{j}(x,y,z)\, dx\, dy\, dz$. Here $\psi_{k}$ is the $k^{\text{th}}$ unperturbed eigenmode of the cavity and the integral is taken over the volume of the membrane.

Applying this theory to the four cavity modes of interest (the singlet and triplet modes) quantitatively reproduces the large-scale ($\sim$GHz) sinusoidal shape of $\omega (x)$ seen in \f{1}{b} if we assume Re[$n$]$ = 2.0$ and $t = 39$ nm (black lines). The discrepancy between the fitted and nominal values of $t$ presumably reflects a combination of fabrication tolerances ($\sim 5$ nm) and the limits of a first-order perturbation theory that includes only four eigenmodes. Nonetheless, the intermediate-scale ($\sim$10 MHz) features in \f{1}{c}(i-ii), such as the lifting of the triplet's degeneracy, are also quantitatively reproduced (\f{1}{c}(iii-iv)). Small-scale features such as avoided crossings agree with the model reasonably well and are discussed below.

\section{Tunability of quadratic coupling}

\begin{figure}
\begin{center}
\includegraphics[width=6.5in]{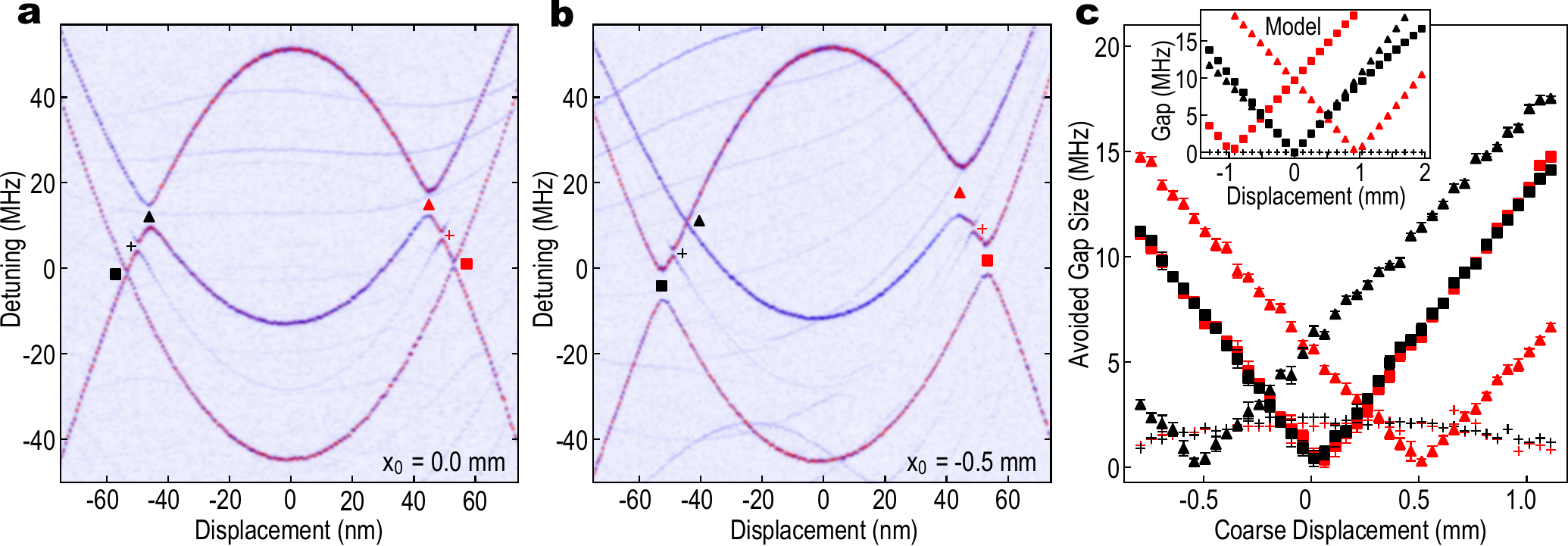}
\end{center}
\caption{Transmission spectrum with the membrane tilted 0.48 mrad about $\tilde{z}$, \textbf{a} positioned near the waist ($x = 0$~$\upmu$m) and \textbf{b} displaced -0.5 mm from the waist. \textbf{c} Magnitudes of the gaps indicated in \textbf{a} \& \textbf{b} plotted versus the membrane's displacement. (inset) The magnitudes of the same gaps, as calculated by the model described in the text.}
\label{fig2}
\end{figure}

The second result of this article is that $\omega^{\prime\prime}$ can be tuned over a wide range by moving the membrane along the $\tilde{x}$ axis. This tunability is important because in some situations it may be desirable to decrease $\omega^{\prime\prime}$ in order to relax other experimental constraints. For example, if $\omega_s$  ($\propto 1/\omega^{\prime\prime}$) is small enough the membrane's motion may result in non-adiabatic transfer of light between the two cavity modes via Landau-Zener-St\"{u}ckelberg-like transitions.\cite{2010_PRAr_Heinrich_Marquardt}

The tunability of $\omega^{\prime\prime}$ is illustrated in \f{2}{a-b}, which each show six avoided crossings between the singlet and triplet modes. When the membrane is at the cavity waist (\f{2}{a}) the upper gaps (triangles) are open and the lower gaps (squares) are closed. When the membrane is displaced 500 $\upmu$m (\f{2}{b}) from the waist, the two lower gaps open, the upper right gap opens further, and the upper left gap closes. The full dependence of $\omega_s$ on membrane position is shown in \f{2}{c}. 

The perturbative model (\f{2}{c} inset) reproduces the linear dependence of $\omega_{s}(x)$ as well as the slope $\partial\omega_{s}/\partial x$ measured for each of the six gaps. The model differs from the data by a constant offset $\sim 3-4$ MHz for the middle and upper gaps (triangles and crosses in \f{2}{c}), which we attribute to asymmetry in the cavity that is not included in the model. We find below (\f{3}{c}) that it is possible to compensate for this intrinsic asymmetry by adjusting the \textit{axis} of the membrane tilt.

\section{Tunable coupling between motion and optical relaxation}

\begin{figure}
\begin{center}
\includegraphics[width=5in]{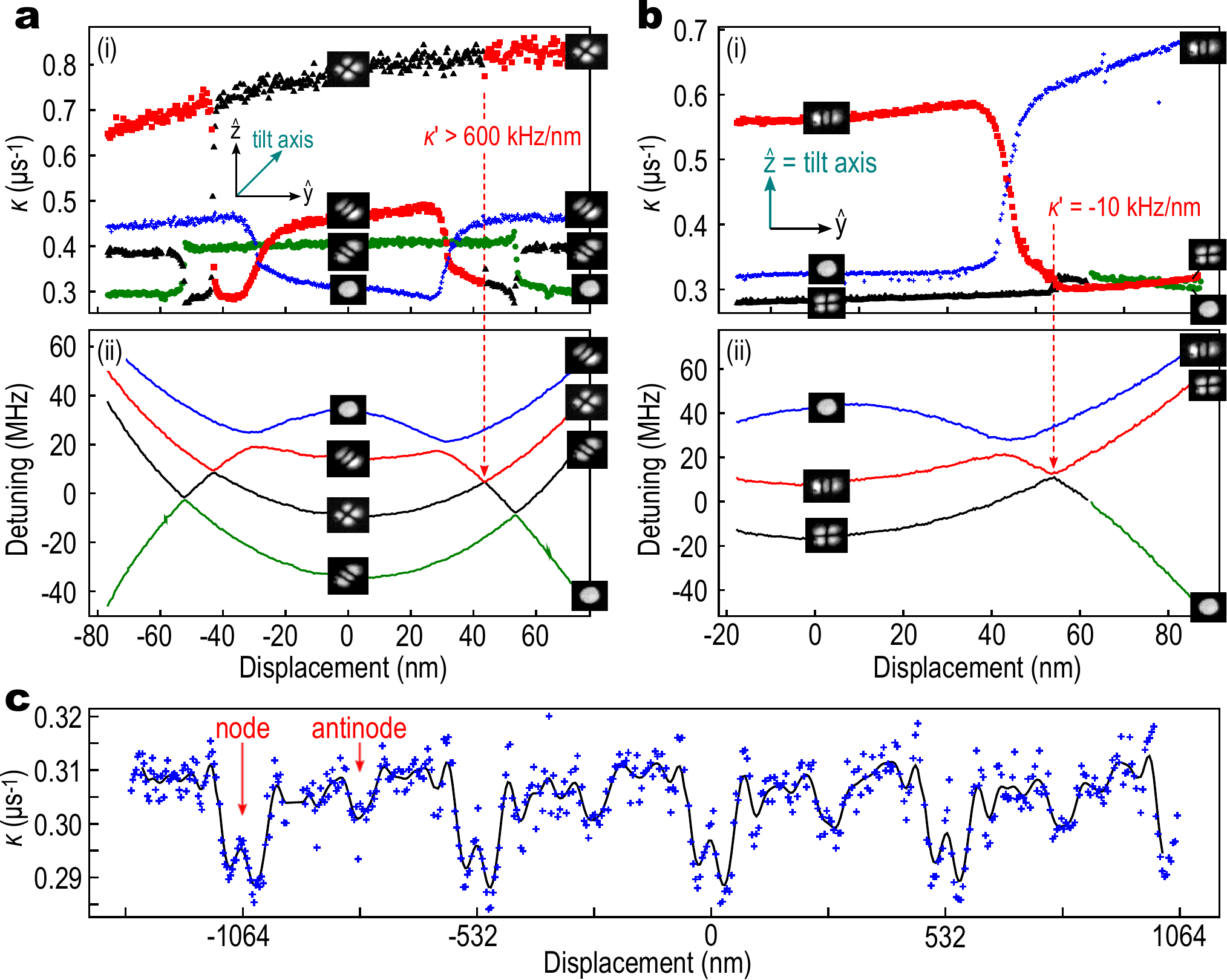}
\end{center}
\caption{\textbf{a} Measurements of (i) cavity relaxation rate and (ii) detuning for each cavity mode versus position for the membrane tilted 0.66 mrad about the axis indicated in the figure. \textbf{b} Same but for the membrane tilted 0.61 mrad about a different axis (indicated in the figure). The dotted arrows highlight the crossing whose gradient $\kappa^\prime$ reverses sign. \textbf{c} Relaxation rate $\kappa$ of the TEM$_{00}$ mode as a function of membrane position. Symbols are data, the black solid line is a smoothed version of the data to show the $\lambda/2 = 532$~nm periodicity.}
\label{fig3}
\end{figure}

In addition to creating large $\omega^{\prime\prime}$, mixing between the cavity modes also causes the cavity loss rate $\kappa$ to vary with $x$. \F{3}{a}(i) shows $\kappa(x)$ for the singlet and triplet modes (the corresponding detunings are shown in \f{3}{a}(ii)). Far from the crossings each cavity mode has a different $\kappa$, reflecting the different overlap between each mode's transverse profile and the mirrors' inhomogeneities. At each avoided crossing the two modes swap their value of $\kappa$, leading to a large $\kappa^{\prime}\equiv \partial\kappa/\partial x$. The dashed line in \f{3}{a} indicates a crossing at which $\kappa^{\prime} > 600$ kHz/nm. 

Note that this data was taken with the membrane's tilt \textit{axis} rotated by 45\degrees to $(\tilde{y} + \tilde{z})/\sqrt{2}$, and the transverse eigenmodes  have followed (inset camera images). The ability to rotate the transverse mode profiles provides some control over which portions of the mirrors the optical modes sample, and enables us to tune $\kappa^{\prime}$ \textit{in situ}. In \f{3}{b} we have rotated the membrane tilt axis back to $\tilde{z}$, with the result that the same crossing shows $\kappa^{\prime} = -10$ kHz/nm. Significantly, the sign of $\kappa^{\prime}$ has changed, implying that at an intermediate tilt axis $\kappa^{\prime} = 0$.

Gradients in $\kappa$ can have several consequences. A linear variation of $\kappa$ with $x$ could preclude a QND measurement of $n_m$, so the ability to tune $\kappa^{\prime}$ to zero is appealing. Separately, it has been predicted that large values of $\kappa^{\prime}$ can be used to laser cool the membrane to its ground state even when the device is not in the resolved sideband regime.\cite{2009_PRL_Elste_Clerk} However numerical estimates show that even with $\kappa^{\prime} = 600$ kHz/nm, the usual optomechanical coupling\cite{2007_PRL_Marquardt_Girvin, 2007_PRL_Rae_Kippenberg} is more promising for laser-cooling the devices described here.

Variations in $\kappa$ can also arise from optical absorption in the membrane. Previous measurements using non-stoichiometric SiN$_x$ membranes found that variations in $\kappa$ were proportional to the overlap of the intracavity standing wave and the membrane, and were consistent with Im$(n) \approx 2 \times 10^{-4}$ for $\lambda = 1064$ nm.\cite{\njp} Subsequent work found lower absorption in stoichiometric Si$_3$N$_4$ membranes, with Im$(n) \lesssim 10^{-5}$ for $\lambda = 985$ nm.\cite{2009_PRL_Wilson_Kimble} \F{3}{c} shows $\kappa$ for the TEM$_{00}$ mode (using a Si$_{3}$N$_{4}$ membrane and $\lambda = 1064$ nm) as $x$ is varied over a few $\lambda$. The small-scale ($\ll \lambda$) variations in $\kappa(x)$ are reproducible and periodic, and presumably arise from mixing with higher-order modes. To estimate the contribution to $\kappa(x)$ from absorption in the membrane (as opposed to mixing with higher-order modes), we extract the Fourier component of $\kappa(x)$ corresponding to the overlap between the membrane and the intracavity standing wave. This sets an upper limit of Im$(n) \lesssim 1.5 \times 10^{-6}$ at $\lambda = 1064$ nm. The lower optical absorption observed here should enable the use of cavities with higher $F$ while decreasing the quantum noise in the cavity and the heating of the membrane.

\section{Feasibility of observing mechanical energy quantization}

To estimate the feasibility of observing energy quantization in the membrane, we first use the expressions derived elsewhere\cite{\nature} for $\Sigma^{(0)}$, the signal-to-noise ratio for measurement of a single quantum jump. Assuming a single-port cavity\cite{2009_PRL_Miao_Chen}, we find $\Sigma^{(0)} = 1$ for a device with the parameters listed in the first row of Table 1. This scheme requires higher finesse and smaller values of $m$ and $\omega_m$ (which could be realized by patterning the membrane into a free standing pad supported by narrow beams), as well as cryogenic pre-cooling to $T=300$~mK. The key advances presented in this paper regarding the realization of this scheme are the demonstration of (i) sufficiently large $\omega^{\prime\prime}$ and (ii) sufficiently low optical loss to meet these requirements. 

\begin{table}
	\centering
		\begin{tabular}{r l l l l l r}
			\hline
			\hline
			& $F$ & $\omega_m/2\pi$ & $m$ & $Q_m$ & $x_0$ &  \\
			\hline
			Quantum Jumps: & 500,000 & 100 kHz & 50 pg & 1.2$\times 10^7$ & -- & $\Sigma^{(0)}$ = 1 \\
			Phonon Shot Noise: & 50,000  & 1 MHz   & 40 ng & 1.2$\times 10^7$ & 2 nm & $\mathcal{S}$ = 20.6\\
			\hline
			\hline
		\end{tabular}
	\caption{Comparison of parameters (cavity finesse $F$, membrane resonance frequency $\omega_m$, mass $m$, and quality factor $Q_m$, mechanically driven to amplitude $x_0$) to observe energy quantization in a mechanical resonator for the two schemes. Both cases assume 5 $\upmu$W of 1064-nm light incident on a 6.313-cm cavity, a bath temperature $T = 300$ mK (laser-cooled $n_{T} < 0.2$), and coupling strength $\omega^{\prime\prime}/2\pi = 4.5$~MHz/nm$^2$.}
	\label{table1}
\end{table}

The requirements for observing phonon shot noise in the membrane are described in Ref.\ [\citeonline{2010_arxiv_Clerk_Harris}]. The ratio between the phonon shot noise signal and the measurement imprecision for such a measurement is $\mathcal{S} = 8 \bar{n}_m n_T \Sigma^{(0)}$, where $n_T = k_{\mathrm{B}} T / \hbar \omega_m$. The parameters listed in the second row of Table 1 result in $\mathcal{S} = 20.6$. Significantly, these parameters correspond to the device demonstrated here; the value of $\omega^{\prime\prime} = 4.5$ MHz/nm$^2$ corresponds to the avoided crossing resolved in \f{1}{d}, and the amplitude of the membrane's motion ($x_0 =2$~nm) is below the onset of dynamical bistability.\cite{2008_APL_Zwickl_Harris} A quality factor $Q_m = 1.2 \times 10^7$ was demonstrated in similar membranes at $T = 300$ mK.\cite{2008_APL_Zwickl_Harris}

\section{Purely-quartic optomechanical coupling}

\begin{figure}
\begin{center}
\includegraphics[width=5in]{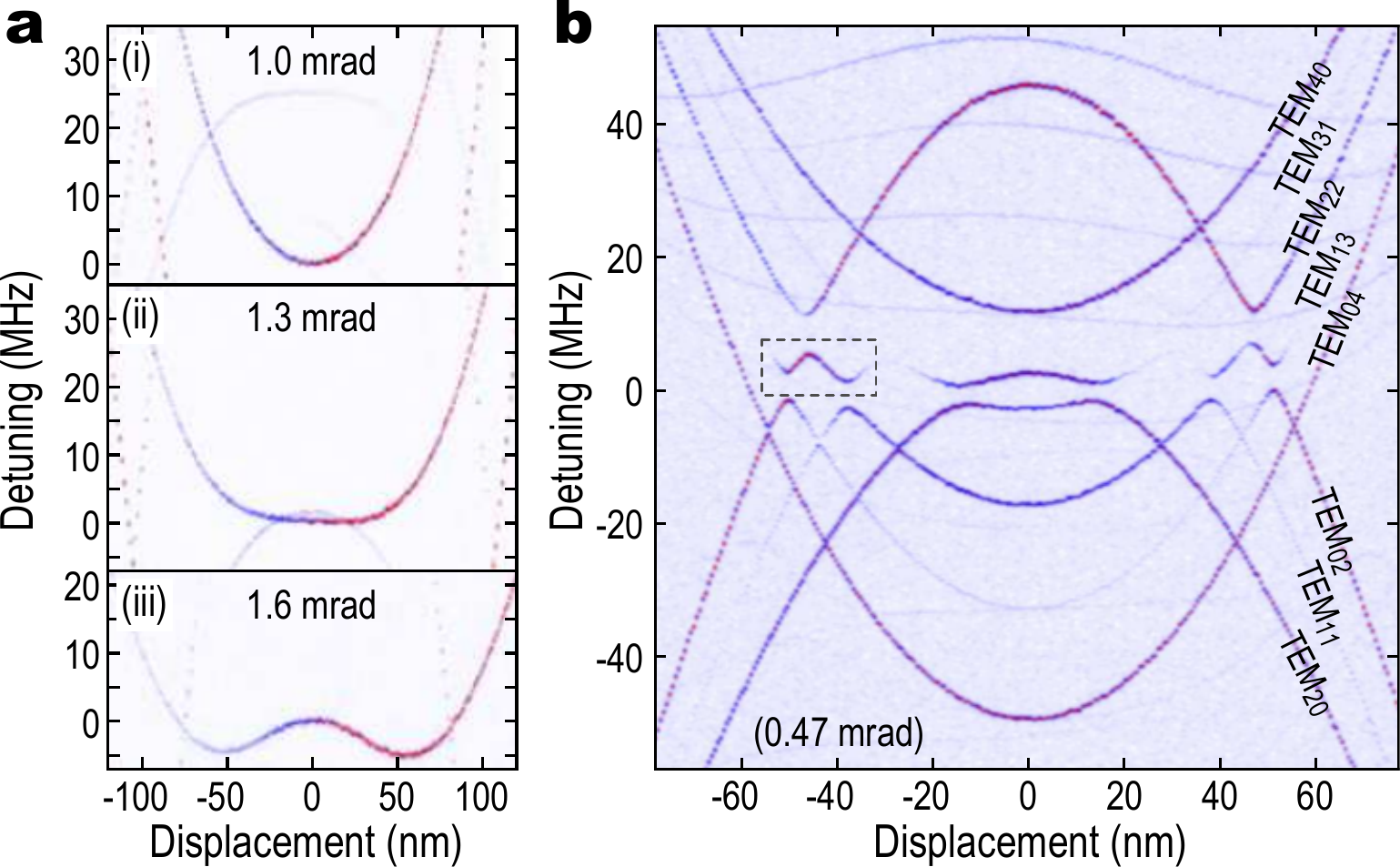}
\end{center}
\caption{\textbf{a} Transmission spectrum for the TEM$_{20}$ mode with the membrane tilted to: (i) 1.0, (ii) 1.3, and (iii) 1.6 mrad, showing transition between quadratic and quartic detuning. \textbf{b} Transmission of the cavity showing the avoided crossings between the TEM$_{\{20,11,02\}}$ and TEM$_{\{40,31,22,13,04\}}$ modes.}
\label{fig4}
\end{figure}

The final point of this article is to demonstrate a new type of optomechanical nonlinearity: quartic ($x^4$) coupling. \F{4}{a} shows that when the membrane tilt is increased to $\sim$ 1 mrad, $\omega(x)$ for the TEM$_{20}$ mode undergoes a smooth transition from $\omega^{\prime\prime} > 0$ (i) to $\omega^{\prime\prime} < 0$ (iii); between these limits $\omega^{\prime\prime} = 0$ (ii) and $\omega (x) \propto x^4$ to lowest order. A similar transition is visible in the faint background modes of \f{2}{a} and \f{4}{b} as a function of mode index. 

This form of $\omega (x)$ can be used to realize an optomechanical coupling described (in the rotating wave approximation) by the Hamiltonian term $H^{(4)}_{\rm{coup}} =$ $\hbar \omega^{(4)} x^{4}_{\rm{zpf}} n_\gamma n_m^2$, with $\omega^{(4)} \equiv \partial^4 \omega^4 / \partial x^4$, $x_{\rm{zpf}} = \sqrt{\hbar/2 m \omega_m}$, and $n_\gamma$ the intracavity photon number. This type of coupling can be used, for example, to prepare Schr\"{o}dinger cat states in the membrane.\cite{2007_PRL_Jacobs}

While the quartic coupling in \f{4}{a} is quite weak ($\omega^{(4)} = 0.4$ Hz/nm$^{4}$), it may be possible to increase $\omega^{(4)}$ using avoided crossings. For example the interaction between the triplet and quintuplet modes (\f{4}{b}) shows avoided crossings in which $\omega^{\prime\prime}$ changes sign. In analogy with the tunability of $\omega^{\prime\prime}$ demonstrated in \f{2}{c}, we expect that careful arrangement of the membrane tilt and position will allow some of the crossings in \f{4}{b} to be purely quartic with a substantially larger $\omega^{(4)}$.

\section{Summary} 

In summary, we have demonstrated an optomechanical device in which the strength and the form of the optomechanical coupling can be tuned over a wide range \textit{in situ}. We have demonstrated control over whether the optical cavity detuning is (to lowest order) linear, quadratic, or quartic in the displacement of a micromechanical membrane, and shown that the quadratic coupling is three orders of magnitude stronger than previously demonstrated. This device also demonstrates extremely low optical loss, and an optical loss gradient that can be tuned to zero. These represent important advances in the ongoing effort to observe and manipulate quantum behavior in a solid mechanical oscillator. 

In particular, the combination of low optical loss and strong quadratic coupling demonstrated here should enable the observation of the membrane's energy quantization without further improvements to the present device. This combination will also enable other functionalities related to quadratic coupling, including dispersive QND readout of the intracavity photon number, two-phonon laser cooling, conditional squeezing between the reflected light and the membrane's motion, and various types of passive optical squeezing.\cite{2008_NJP_Jayich_Harris, NunnenkampInPrep}

\section{Methods}

All measurements were performed at room temperature and pressure $< 10^{-5}$ Torr. The end mirrors were clamped to an invar spacer, and a mount equipped with three motorized actuators (for tilting and displacing the membrane) were mounted to this spacer. Two small piezoelectric elements provided finer positioning of the membrane along the cavity axis. Spectroscopy was performed by sweeping the laser frequency from low to high and stepping the membrane position with the piezo elements. The cavity's optical loss was measured via cavity ringdown.

\section{Acknowledgments}

We would like to thank H. Cao, A. Clerk, F. Marquardt, S. M. Girvin, A. Nunnenkamp, and D. Schuster. This work has been supported by grants from the NSF (\#0855455 and \#0653377) and AFOSR (\#FA9550-90-1-0484). J.G.E.H. acknowledges support from the Alfred P. Sloan Foundation. This material is based upon work supported by DARPA under Award No. N66001-09-1-2100.

\section{Author Contributions}

J.C.S. performed the measurements, developed the perturbation theory, and carried out the data analysis. C.Y., B.M.Z., A.M.J. assisted with each phase of the project. J.G.E.H. supervised each phase of the project.


\end{document}